\documentstyle[prl,aps,epsf,floats,twocolumn]{revtex}

\newcommand{\bA}{{\bf{A}}}

\newcommand{\vrr}{{\vec{r}}}

\newcommand{\vq}{{\vec{q}}}

\def\lst2{{(l^*)^2}}
\def\hz{{\hat z}}

\def\Lam{\Lambda}
\def\dd{d^{\dagger}}

\newcommand{\etab}{\mbox{\boldmath $\eta $}}
\newcommand{\Pib}{\mbox{\boldmath $\Pi $}}
\newcommand{\br}{{\bf r}}
\newcommand{\bp}{{\bf p}}
\newcommand{\bR}{{\bf R}}
\newcommand{\bq}{{\bf q}}

\newcommand{\barro}{\bar{\bar{\rho}}({\bf q})}
\newcommand{\barchi}{\bar{\bar{ \chi}}({\bf q})}

\newcommand{\eeq}{\end{equation}}

\newcommand{\beq}{\begin{equation}}

\def\t{{\theta}}
\def\ihalf{{i\over2}}
\def\half{{1\over2}}
\def\third{{1\over3}}

\def\bbrho{{\bar{\bar \rho}}}
\def\intq{{\int {d^2q\over(2\pi)^2}}}

\def\eqa{\begin{eqnarray}}
\def\eea{\end{eqnarray}}
\def\a{{\alpha}}

\def\prl{{Phys. Rev. Lett.\ }}
\def\prb{{Phys. Rev. B\ }}
\begin{document}
\draft
\flushbottom
\twocolumn[
\hsize\textwidth\columnwidth\hsize\csname @twocolumnfalse\endcsname
\title{ Hamiltonian Theory of the Fractional Quantum Hall Effect: Effect of Landau Level Mixing }
\author{  G. Murthy and   R.Shankar  }
\address{Department of Physics and Astronomy, University of Kentucky, Lexington KY 40506-0055,\\
 Department of Physics, Yale University, New Haven CT 06520}
\date{\today}
\maketitle

\begin{abstract}
 We derive an effective hamiltonian in the Lowest Landau Level (LLL)
 that incorporates the effects of Landau-level mixing to all higher
 Landau levels to leading order in the ratio of interaction energy to
 the cyclotron energy, $\kappa = (e^2/\varepsilon l)/\omega_c$. We
 then transcribe the hamiltonian to the composite fermion basis using
 our hamiltonian approach and compute the effect of LL mixing on
 transport gaps.
\end{abstract}
\vskip 1cm
]

\section{Introduction}

The fractional quantum Hall effects (FQHE)\cite{fqhe-ex,perspectives}
have demonstrated the existence of new states of matter for
two-dimensional electron gases (2DEG's) in high magnetic fields. When
the fields are high enough that only a fraction of a Landau level (LL)
is occupied, the system reorganizes itself at special fractions into
incompressible ground states\cite{laugh} with fractionally charged
excitations. The current understanding of the FQHE is schematically
expressed by saying that the quasiparticles are electrons dressed by
an even number of units of statistical flux, called Composite Fermions
(CF's)\cite{jain-cf,jain-cf-review}. The FQHE of electrons is then
understood as the integer quantum Hall effect of CF's.

A fruitful approximation in the study of the FQHE has been the
restriction of the single particle electronic wavefunctions to the
LLL. This approximation becomes exact in the limit of very high
magnetic fields ($\kappa = (e^2/\varepsilon l)/\omega_c \to 0$) and
captures the essential physics underlying the FQHE. However, when one
wishes to make detailed comparisons between the theory and experiment,
say in the study of transport gaps, one must seek corrections due to
the nonzero value of $\kappa$, that is, include the effects of Landau
level mixing. Hitherto these effects have been studied 
numerically\cite{yoshioka,price,melik1}, with the latest results
coming from a fixed-phase diffusion Monte Carlo study\cite{melik2}.

In this paper we present an analytical calculation of the effect of
LL-mixing in two steps. First we deduce an effective hamiltonian
within the LLL which incorporates the leading contribution in $\kappa$
due to higher LL's. This hamiltonian is written in terms of electronic
coordinates. We then invoke results from our previous work on the
hamiltonian formulation of the FQHE\cite{us1,us2,aftermath} which
allows one to rewrite the hamiltonian in terms of composite fermion
degrees of freedom. The advantage is that these particles see an
effective field which is just right to fill an integer number of LL's,
thus yielding a nondegenerate ground state on which we can base a
Hartree-Fock calculation of gaps.

Specifically, the hamiltonian approach provides an expression for the
LLL projected electron density $\bar{\bar{\rho}}(\bq )$ in the CF
basis\cite{us1,us2,aftermath}. Since the effective hamiltonian is
expressed in terms $\bar{\bar{\rho}}(\bq )$, this all we need.

In Section II, we derive the effective hamiltonian and in Section III
we express it in the CF basis. Section IV contains the calculation of
gaps and conclusions follow in Section V. Details are  relegated
to the two Appendices. While this paper focusses on the calculation of
transport gaps, the approach presented here can be extended to
calculations of other physical quantities in the Hamiltonian theory\cite{rest}.

\section{The effective hamiltonian-electronic basis}

Our starting point will be the full electronic Hamiltonian, which is given by 

\begin{eqnarray}
H&=&\sum_{\alpha}a^{\dag}_{\alpha}a_{\alpha} E_i + {1\over 2} \int {d^2q\over
2\pi^2}\rho_{12}(\bq )\rho_{34}(-\bq ) v(q)a^{\dag}_{1}
a^{\dag}_{3}a^{}_{4}a^{}_{2}\\ & \equiv &H_0 +V
\end{eqnarray}

where any subscripts $\alpha$ stands for pair of indices $(n_{\alpha},
m_{\alpha})$ ($n$ labels the LL index and $m$ the angular momentum
within the LL). Both labels assume integral values from zero to
infinity and

\beq E_{\alpha} = n_{\alpha}\omega_0 
\eeq 

with $\omega_0$ being the cyclotron frequency. We have dropped the
zero-point energy and kept only the normal-ordered part of the
interaction. The operators $a$ and $a^{\dag}$ are electron destruction
and creation operators, $v(\bq )$ is the electron-electron interaction 
and $\rho_{\alpha \beta}(\bq )$ are the matrix elements of the charge
density operator. These matrix elements can be determined using the
standard decomposition of the electron coordinates and momenta into
cyclotron ($\etab^e$) and guiding center (${\bf R}^e$) variables

\eqa
\etab_e=&l^2\hz\times\Pib_e\\
\bR_e=&\br_e- l^2\hz\times\Pib_e
\eea

where $\Pib_e=\bp_e+e\bA$ is the velocity operator of the electrons and
$l= \sqrt{1/eB}$ is the magnetic length. These coordinates obey the
commutation relations

\beq 
\left[ \etab_{ex} \ , \etab_{ey} \right] = il^2 \ \ \
\left[ \bR_{ex} \ , \bR_{ey} \right]=-il^2 \ \ \  \ \left[ \etab_e \ , \bR_e
\right]=0 
\eeq 

Expressing the electron coordinate as $\br_e=\bR_e+\etab_e$ we have

\begin{eqnarray}
\rho_{\alpha \beta}(\bq ) &=& \langle n_{\alpha}  m_{\alpha}
|e^{-i\bq \cdot ({\bf R}_e+\etab_e })|n_{\beta} m_{\beta} \rangle\\ &=&
\langle m_{\alpha} |e^{-i\bq \cdot {\bf R}_e }| m_{\beta} \rangle
\times \langle n_{\alpha}
 |e^{-i\bq \cdot \etab_e }|n_{\beta}  \rangle \\
 &=& \rho_{\alpha \beta}^{(m)}\times \rho_{\alpha \beta}^{(n)}
\end{eqnarray}

where the superscripts $m$ and $n$ indicate that the matrix elements
correspond to the guiding center and cyclotron coordinates
respectively. These matrix elements can be expressed in terms of
Laguerre polynomials. However, for this section we only need to know
that

\begin{eqnarray}
\rho^{(n)}_{00}(\bq )&=& e^{-q^2l^2/4}\\ \sum_{\beta}
\rho^{(m)}_{\alpha \beta }(\bq_1 ) \rho^{(m)}_{ \beta
\gamma}(\bq_2 )&=&\rho^{(m)}_{\alpha \gamma }(\bq_1 +\bq_2)\
e^{i\bq_1 \times \bq_2 l^2/2}\label{mta}
\end{eqnarray}

which reflects the magnetic translation algebra: 

\beq e^{-i \bq_1
\cdot \bR_e} e^{-i \bq_2 \cdot \bR_e} =e^{-i (\bq_1 +\bq_2 )\cdot \bR_e}
e^{i(\bq_1 \times \bq_2 ) l^2/2}  
\eeq

It will prove convenient to define

\beq 
\int_q = \int{d^2 q\over 4\pi^2} v(q). \label{measure}
\eeq
where the potential is included as part of the measure. 

The leading term in the  effective hamiltonian is the  obvious
contribution from $H_{00}$,  the restriction of $H$ to the LLL:

\eqa
H_{00} &=& \sum_m a^{\dag}_{0m}a_{0m} \cdot  0 \cdot\omega_0\nonumber 
\\ & & +{1 \over 2} \int_q \left[ a^{\dag}_{0m_1}
a^{\dag}_{0m_3}  a_{0m_4}a_{0m_2} \right. \nonumber \\ & &\left. \cdot
 \rho_{m_1m_2}^{(m)}(\bq )\rho_{m_3m_4}^{(m)}(-\bq
)\rho_{00}^{(n)} (\bq ) \rho^{(n)}_{00}(-\bq )\right] \nonumber \\
&=& {1 \over 2} \int_q a^{\dag}_{m_1} a^{\dag}_{m_3} a_{m_4} a_{m_2}
 \rho_{m_1m_2}^{(m)}(\bq )\rho_{m_3m_4}^{(m)}(-\bq
)e^{-q^2l^2/2}\label{0}
\eea

where we have  labeled LLL operators with just an $m$ index. In
terms of $\barro $, the density restricted to the LLL and defined
by 
\beq 
\barro  = \sum_{ij} a^{\dag}_{m_i}a_{m_j}\rho_{m_im_j}
(\bq ) 
\eeq 
we can write 
\beq 
H_{00}= V={1\over 2} \int_q e^{-(ql)^2/2}\barro
\bar{\bar{\rho}}(-\bq ). 
\eeq

Here we have removed the normal-ordering, which leads to an
unimportant additive constant in the energy.  The operator $\barro$
has the same commutators as $e^{i\bq \cdot
\bR}$.

To extract the leading correction due to higher LL's we write the
Schr\"{o}dinger equation as 

\beq 
\left[
\begin{array}{cc}
  H_{00} & H_{0n'} \\
  H_{n'0} & H_{n'n'}
\end{array} \right]   \left[ \begin{array}{c} \phi \\ \chi
\end{array} \right] = E \left[ \begin{array}{c} \phi \\ \chi
\end{array} \right]
\eeq 

where $\phi$ is  restricted to the space spanned by Fock
states composed  of just the LLL states and $\chi$ stands for
everything else. The exact equation obeyed by $\phi$ is 

\beq
\left( H_{00} + H_{0n'}{1 \over E-H_{n'n'}}H_{n'0}\right) \phi = E
\phi 
\eeq 

which is not an eigenvalue problem since $E$ appears on
both sides. However we may approximate as follows: 

\beq {1 \over
E-H_{n'n'}}=-{1 \over H_{n'n'}} + {\cal O} (v/\omega)
\eeq

since the eigenvalue $E$ we are interested in is of order $v$ and
 the eigenvalues of $H_{n'n'}$ are of order $\omega_0$. To the
 same accuracy
 in $\kappa = v/\omega_0$ we can also replace

\beq
 H_{n'n'}\simeq  H_{n'n'}^{0} = \sum_{\alpha}
 a^{\dag}_{\alpha}a_{\alpha} n^{'}_{\alpha} \omega_0
\eeq

In fact, one can do a little better than this, and incorporate the
interactions into $H_{n'n'}$ in by including the Hartree-Fock energy
of the $n'^{th}$ level. This can be viewed as a perturbation theory in
the {\em non-HF part} of the interaction. We will present results both
with and without this HF energy in $H_{n'n'}$. To maintain generality
we define a reduced energy
\beq
\varepsilon(n)=n+{\nu\over \omega_0} \intq v(q) (|\rho^{(n)}_{00}(\vq)|^2-
|\rho^{(n)}_{n0}(\vq)|^2)
\label{ehf}\eeq
where we have used the fact the the FQH states have uniform
occupations of $\nu$ in the LLL, and we have normalized the LLL energy
to be zero.  This leads us to the effective Hamiltonian

\beq
 H^{eff}_{00}=H_{00} +\delta H_{00} 
\eeq

where

\begin{eqnarray} \delta H &=& -H_{0n'}{1\over H^{0}_{n'n'}}H_{n'0}\\
 &=&-{1\over 4}\int_{q_1}\int_{q_2}\left[ \sum^{'}_{1\ldots
8}a^{\dag}_{8}
a^{\dag}_{6}a^{}_{5}a^{}_{7}{1\over H_{n'n'}^{0}} a^{\dag}_{1}
a^{\dag}_{3}a^{}_{4}a^{}_{2}\right.\nonumber \\ &&\left.  
\rho_{12}(\bq_1 )\rho_{34}(-\bq_1 ) \rho_{87}(\bq_2
)\rho_{65}(-\bq_2 )\right]
\end{eqnarray}

The prime on $\sum'$ denotes the fact that if $\delta H_{00} $ is
to act on the LLL sector, we need 

\beq 
n_8=n_6=n_4=n_2=0.
\eeq

This means we can have two kinds of terms:

\begin{itemize}
\item $\delta H_{00}^{(2)}$, a  term in which both $a_{1}^{\dag}$ and
$a_{3}^{\dag}$
excite the electron to higher LL and $a_7$ and $a_5$ bring it back
to the LLL.
\item $\delta H_{00}^{(1)}$ in which either one of  $a_{1}^{\dag}$ or 
$a_{3}^{\dag}$
excites the electron to higher LL and one of $a_7$ or $a_5$
brings it back to the LLL.
\end{itemize}

It is readily seen that 

\begin{eqnarray} 
\delta H_{00}^{(2)} &=&
-{1\over 4} \int_{q_1}\int_{q_2}\left[ \sum_{1\ldots
8}\rho_{12}(\bq_1 )\rho_{34}(-\bq_1 ) \rho_{87}(\bq_2
)\rho_{65}(-\bq_2 )\right.  \nonumber \\ &&  \left. a_{8}^{\dag}
a_{6}^{\dag}a_4a_2
 {1 \over
(\varepsilon(n'_1)+\varepsilon(n'_3))\omega_0}(\delta_{53}\delta_{71}-\delta_{51}\delta_{73})\right]
\end{eqnarray}
The contribution from the first delta function is

\begin{eqnarray}
{}  &&\!\!\!\!\!\!\!\!\!\!\!\!\!\!\!\!\!\!\!\!\!
\!\!\!\!\!\!\!\!\!\!\!\! \!\!{-1\over 4\omega_0 }
\int_{q_1q_2}\!\! \left[ \sum_{m} \rho_{m_8m_1}^{(m)}(\bq_2
)\rho_{m_6m_3}^{(m)}(-\bq_2 )
\rho_{m_1m_2}^{(m)}(\bq_1 )\rho_{m_3m_4}^{(m)}(-\bq_1 ) \right. \nonumber
\\ &&  \left. F_1(q_1,q_2)a^{\dag}_{0,m_8}
a^{\dag}_{0,m_6}a^{}_{0,m_4}a^{}_{0,m_2}\right] \nonumber \\
F_1(q_1,q_2)&=&\!\!\! \sum_{n_1,n_3=1}^{\infty}\!\!\!
{\rho_{0,n_1}^{(n)}(\bq_2)
\rho_{n_1,0}^{(n)}(\bq_1)\rho_{0,n_3}^{(n)}(-\bq_2)
\rho_{n_3,0}^{(n)}(-\bq_1)\over \varepsilon(n_1)+\varepsilon(n_3)}
\end{eqnarray}

Using completeness to do the sum over $m_1$ and $m_3$, using the
magnetic algebra Eqn. (\ref{mta} ) and doubling the answer because
of the other delta function which makes an equal contribution
(upon relabeling of dummy indices) we arrive at

\begin{eqnarray}
\delta H_{00}^{(2)} &=& {-1\over 2\omega_0}\int_{q_1q_2}\!\!\!\!
e^{-i{\bq_1 \times \bq_2 l^2}} \!  \bar{\bar{\rho}} (\bq_1
+ \bq_2 )\bar{\bar{\rho}} (-\bq_1\! - \bq_2 )F_1(q_1,q_2)\nonumber
\\
& &  \label{2body}
\end{eqnarray}

Now for the term with just one higher LL
excitation:

\begin{eqnarray}
\delta H_{00}^{(1)} &=& -{1\over 4} \int_{q_1}\int_{q_2}\left[
\rho_{12}(\bq_1 )\rho_{34}(-\bq_1 ) \rho_{87}(\bq_2
)\rho_{65}(-\bq_2 ) \right. \nonumber \\ &&\left. \!\!\! \!\!\!\!\!\!
\!\!\!\!\!\!\!\!\!\!\!\! a_{8}^{\dag}
a_{6}^{\dag}a_5a_7{1\over H_{n'n'}^{0}}a_{1}^{\dag}
a_{3}^{\dag}a_4a_2 \right]
\end{eqnarray}

Consider the contribution when $a_{1}^{\dag}$ takes the electron
up to a higher LL and  and $a_7$ brings it down. The other three
possibilities obtained by permuting $1\longleftrightarrow 3$ and
$5 \longleftrightarrow 7$ make the same contribution upon
relabeling of dummy indices. The net contribution due to these
terms in which a single excitation to a higher LL occurs is

\begin{eqnarray}
\delta H_{00}^{(1)} &=& - \int_{q_1}\int_{q_2}\sum_{1\ldots
8}^{'}a_{m_8}^{\dag}
a_{m_6}^{\dag}a_{m_5}a_{m_7}{\delta_{71}}a_{m_1}^{\dag}
a_{m_3}^{\dag}a_{m_4}a_{m_2}\nonumber \\ && \left[
{\rho_{81}(\bq_2 )\rho_{12}(\bq_1 )\over \varepsilon(n_1) \omega_0}\right]
\rho_{34}(-\bq_1 )\rho_{65}(-\bq_2 )
\end{eqnarray}

\begin{eqnarray}  &=& -\int_{q_1}\int_{q_2}\left[\sum_{1\ldots
8}
\rho_{m_8m_1}(\bq_2 )\rho_{m_1m_2}(\bq_1 )\times\right. 
\nonumber \\ & &\left. 
\!\!\!\!\!\!\!\!\!\!\!\!\!\!\!\!\!\!\!\!\!\!\!\!\!\!\!\!\!\!\!\!
G(\bq_1 , \bq_2 )\sum_{m}\rho^{(n)}_{00}(-\bq_1 ) \rho^{(m)}_{m_3m_4}(-\bq_1
)\rho^{(n)}_{00}(-\bq_2 ) \rho^{(m)}_{m_6m_5}(-\bq_2 )\right. \nonumber \\
&& \left. 
a_{m_8}^{\dag}a_{m_6}^{\dag}a_{m_5}a_{m_3}^{\dag}a_{m_4}a_{m_2}\right]
\\ &=&-\int_{q_1}\int_{q_2}\left[ e^{-i(\bq_1 \times \bq_2
)l^2/2}e^{-(q_{1}^{2}+ q_{2}^{2})l^2/4} \times 
\right. \nonumber
\\ &&\left. 
\!\!\!\!\!\!\!\!\!\!\!\!\!\!\!\!\!\!\!\!\!\!\!\!\!\!\!\!\!\!\!\!
G(\bq_1 , \bq_2 )\sum_{m}\rho_{m_8m_2}(\bq_1 +\bq_2 )\rho_{m_3m_4}(-\bq_1
)\rho_{m_6m_5}(-\bq_2 ) \right. \nonumber \\ && \left. 
a_{m_8}^{\dag}a_{m_6}^{\dag}a_{m_5}a_{m_3}^{\dag}a_{m_4}a_{m_2}\right]
\\ G(\bq_1 , \bq_2 ) &=&\sum_{n'>0}{\rho^{(n)}_{0n'}(\bq_2 )\rho
^{(n)}_{n'0}(\bq_1 )\over \varepsilon(n')\omega_0 }
\end{eqnarray}

If we now rearrange the operators (whose subscripts now all
describe $m$) so as to match those on $\rho_{ij}$ as follows

\eqa
a_{8}^{\dag}a_{6}^{\dag}a_{5}a_{3}^{\dag}a_{4}a_{2}&=&\bigg(
a_{8}^{\dag}a_{2}a_{6}^{\dag}a_{5}^{\dag}a_{3}^{\dag}a_{4}
\nonumber
\\ &-& \delta_{32}a_{8}^{\dag}a_{6}^{\dag}a_{5}a_{4}\nonumber \\
&-& \delta_{62}a_{8}^{\dag}a_{5}a_{3}^{\dag}a_{4}\bigg)
\eea

we arrive at the following expression for $\delta H_{00}^{(1)}$

\begin{eqnarray}
\delta H_{00}^{(1)}&=& -\int_{q_1}\int_{q_2}  
G(\bq_1,\bq_2)e^{-(q_{1}^{2}+ q_{2}^{2})l^2/4}\times
\nonumber
\\ &&  \left[e^{-i(\bq_1\times \bq_2 )l^2/2} \bbrho(\bq_1+\bq_2 )  \bbrho(-\bq_2)
 \bbrho(-\bq_1 ) \right. \nonumber  \\
 && \left. \!\!\!\!\!\! -\bbrho(\bq_2)\bbrho(-\bq_2)- e^{-i(\bq_1
\times \bq_2 )l^2}\bbrho(\bq_1 )
\bbrho(-\bq_1  )\right]\label{3body}
\end{eqnarray}

In writing the answer in terms of $\bbrho$'s we must pay attention to
the order since they do not commute with each other. It should also be
noted that the first term of Eqn. (\ref{3body}) produces a
contribution to another effective two-body term via
\beq
\bar{\bar{\rho}}(\bq_1+ \bq_2 ) ={\nu\over 2\pi l^2} (2\pi)^2 \delta^2(\bq_1+\bq_2) + :\bar{\bar{\rho}}(\bq_1+ \bq_2 ):
\eeq
where the first term contains the density $\nu/2\pi l^2$.

The final effective hamiltonian is given by 
\beq 
H_{00}^{eff}=H_{00}+\delta H_{00}^{(2)}+\delta  H_{00}^{(1)} \label{heff}
\eeq

where $ H_{00}$, $\delta H_{00}^{(2)}$ and $\delta H_{00}^{(1)} $
are given by Eqns.(\ref{0}), (\ref{2body}) and (\ref{3body}).

\section{The effective hamiltonian in the CF basis}

The preceding calculation of the effective LLL theory in terms of
electrons could have been performed decades ago since it is not
dependent on any particular representation of $\bbrho$. However,
written in terms of electronic variables, the effective theory suffers
from the same handicap as the original one: the noninteracting part of
the hamiltonian has a huge ground state degeneracy for filling
fraction $\nu ={p\over 2ps+1} <1$. To get a nondegenerate starting
point, we must transcribe $H_{eff}(\bar{\bar{\rho}}(q))$, calculated
above in the electronic basis, to the CF basis. The CF description
will be in terms of particles that see a weaker field of just the
right strength that they fill exactly $p$ LL's, with no degeneracy in
the ground state. Over the years we have developed a route that takes
one from the electron-representation to the CF-representation. We will
not describe it here since it is long and has been described
before\cite{us1,us2,aftermath}.  All we need here are the end results,
which look a lot more attractive than the path that led to them,
fraught as it was with approximations and inspired guesses. Instead,
we will arrive at our final result armed with hindsight.

Consider a CF Hilbert space where each fermion is described by a
coordinate $\br$ and momentum $\bp$. (Hereafter, all unsubscripted
variables will represent the CF degrees of freedom.)  From these we
construct the velocity operator
\beq
 {\bf \Pi }={\bf p} +e{\bf A}^*\ \ \ \ \   A^* = {A \over
 2ps+1}
\eeq

 where the weakened vector potential $A^*$ is what the CF sees.
 In terms of these variables, the electron guiding center $\bR_e$ takes
 the form

\begin{eqnarray}
{\bf R}_e &=& {\bf r} -{l^2\over (1+c)}\hat{\bf z}\times {\bf \Pi
}\label{re}, \\ \left[ R_{ex}\ , R_{ey} \right] &=& - il^2,\\ c^2
&=& {2ps\over 2ps+1}
\end{eqnarray}

{\em Note that $\bR_e$ obeys the right commutation relations. Also,
since it is written in terms of an object that sees a weaker field
$A^*=A/(2ps+1)$, if we express $H^{eff}$ in terms of these variables,
we will not encounter the degeneracy problem at the Jain fractions.}
This is the motivation for switching to the new coordinates.

It is useful to write $\bR_e$ in terms of the CF guiding center
and cyclotron coordinates $ \bR$ and $\etab$: 
\beq 
\bR_e= \bR+ c\etab .
\eeq

Since we have embedded $\bR_e$ in a regular fermionic Hilbert
space, we have room for another pair of guiding center-like
coordinates. Let us call them $\bR_v$. These will be of the form

\beq 
\bR_v = \alpha \bR + \beta \etab 
\eeq

Demanding that $ \bR_v$ commutes with $ \bR_e$ gives us $\beta =
\alpha / c$. How about the overall scale of the operator? Here is
where our previous work tells us to choose $\alpha =1$, that is

\beq 
\bR_v =  \bR +  \etab /c.
\eeq 

In terms of $\br$ and ${\bf \Pi
}$

\begin{eqnarray}
{\bf R}_v &=& {\bf r} +{l^2\over c(1+c)}\hat{\bf z}\times {\bf \Pi
}\label{rv}\\ \left[ R_{vx}\ , R_{vy} \right] &=&  il^2/c^2, \ \
\\
 \left[ {\bf R}_e\ , {\bf R}_{v} \right]&=& 0.
\end{eqnarray}

The merit of this choice is the following.  These commutation
relations correspond to the guiding center coordinates of a particle
of charge $-c^2 = -2ps/(2ps+1)$. This is precisely the charge of an
object that must pair with the electron to form the CF and we refer to
it as the {\em pseudovortex} coordinate, since it has the same charge
as a 2s-fold vortex in Laughlin states. It must be emphasized that
$R_v$ cannot be directly identified with the {\em physical} vortex
that exists around an electron. There is, however, a connection
between $\bR_v$ and the physical vortex which arises upon the choice
of a suitable state in the $\bR_v$ variables, as explained below.

$\bR_v$ is a cyclic coordinate that does not enter the hamiltonian,
and so has no dynamics. 
The eigenfunctions of $H$ will be of the form 

\beq 
\Psi
(R_{ex},R_{vx})=\Psi (R_{ex})\Psi (R_{vx})\label{factorize}
\eeq

where we have chosen as a commuting pair of coordinates $R_{ex}$ and
$R_{vx}$ with conjugate momenta $R_{ey}$ and $R_{vy}$. (One can also
use the Bargman representation and consider $\Psi(z,w)$, where $z$ and
$w$ are complex numbers, as we will find convenient to do later.)
Whereas $\Psi (R_{ex})$ is an eigenfunction of $H(R_e)$, $\Psi
(R_{vx})$ is completely arbitrary. Thus each eigenfunction is
infinitely degenerate. This is exactly the kind of degeneracy a gauge
symmetry would introduce. We must therefore ``fix our gauge'' i.e.,
choose an arbitrary function $\Psi_0 (R_{vx})$ to accompany $\Psi
(R_{ex})$.  No physical observable (function of $R_e$) will depend on
this choice.  Our previous derivation based on canonical
transformations\cite{us1,us2,aftermath} naturally led to the following
constraint on $\Psi (R_{vx})$

\beq
\bigg(\sum_je^{-i{\bf q\cdot R_{vj}}}\bigg) |\mbox{physical states}\rangle=0\label{const} 
\eeq

The constraint arose because we had introduced additional
oscillators at the cyclotron scale and these had to be paid for.

In hindsight we see that we are free to take an axiomatic view of
$\bR_e$ and $\bR_v$ and to simply introduce them as the above
functions of CF coordinates and momenta (Eqs.
(\ref{re}-\ref{rv})). The hamiltonian depends on $\bR_e $ alone.  The
constraint we were led to in our earlier work, Eqn.  (\ref{const}), is
an acceptable choice for fixing the gauge.  However, since all
physical observables depend on just $\bR_e$, and $\bR_e$ commutes with
$\bR_v$, {we may assign to $\bR_v$ any dynamics we want without
changing the physics}.  For example, we could add to $H (\bR_e )$ a
piece $H(\bR_v )$ which is any generic repulsive interaction and
demand that $\Psi_0 (R_{vx})$ be its ground state. Now it turns out
that for any generic repulsion there is essentially a unique answer,
the $\nu=1/2s$ bosonic Laughlin wavefunction\cite{dh}. The reason is
as follows. First, we must choose $R_v$ to be a bosonic coordinate
(the wavefunctions have to be antisymmetric in $\br_e$ and $\bR_e$,
and hence symmetric in $\bR_v$). Next, since $\bR_v$ has a magnetic
algebra charge of $-c^2$, and there is one pseudovortex per electron,
it is easily seen that it is always at filling

\beq 
\nu'= {\nu\over -c^2} = -{p\over 2ps +1}{2ps+1
\over 2ps}=-{1\over 2s}.
\eeq 

which leads us to the wavefunction
\beq
\Psi_{v}^{L}=\prod\limits_{i<j}(z_i-z_j)^{2s} e^{-\sum\limits_{i}^{} {c^2|z_i|^2\over4}}
\label{half-bos-laugh}\eeq
Note that the magnetic length appropriate to the bosons (with charge
$c^2e$) has been used.

Now it might seem that all these virtues are moot since $\bR_v$ is a
fictitious coordinate whose dynamics should make no difference to the
actual physics. While this is certainly the case in any exact
calculation, in the approximate HF calculations we will employ (where
the wavefunction does not have the factorized form of
Eqn. (\ref{factorize}), different choices give different answers and
the above choice gives the wave functions with the best correlations.

To see this, let us first define the CFHF wavefunction. Here we work
in the CF single-particle basis in the effective field $B^*$ (defined
in more detail in the next section), and fill the lowest
single-particle states upto a filling of $\nu^*=\nu/(2s\nu-1)$, which
reduces to $\nu^*=p$ for $\nu=p/(2ps+1)$. It is easy to
show\cite{rest} that these states are indeed HF states of the
Hamiltonian.

As mentioned above, the CFHF wavefunction cannot be a true eigenstate,
since it not of the required product form of
Eqn. (\ref{factorize}). However, we can ``project'' it in the
following way
\eqa
&{\cal P} \Psi_{CFHF}(\{\bR_{e,i},\bR_{v,i}\})=\\
&|\Psi_v(\{\bR_{v,i}\})><\Psi_v(\{\bR_{v,i}\})|\Psi_{CFHF}(\{\bR_{e,i},\bR_{v,i}\})>\\
&=\Psi_v(\{\bR_{v,i}\})\Psi_{e,\Psi_v}(\{\bR_{e,i}\})
\eea
The result is a wavefunction which is explicitly in the form of a
product. Clearly, different choices of $\Psi_v$ will lead to different
results for the electronic part of the wavefunction, and this
dependence is indicated by the subscript on the electronic
wavefunction. For a given CFHF wavefunction, say that of the ground
state, we can imagine systematically varying $\Psi_v$ to obtain the
electronic wavefunction with the best possible energy. It would then
be reasonable to call this $\Psi_v$ the best possible pseudovortex
wavefunction.

In Appendix A we present the details of how to carry out the
``projection''. The result can be gleaned from an identity which we
borrow from Appendix A;
\beq
\int d^2z_v f(\{{\bar z}_{v,i}\}) e^{-\sum\limits_{k} {c^2|z_{v,k}|^2\over2l^2} +\sum\limits_{k} {c^2z_{v,k}z_{e,k}\over2l^2}}=f(\{{z}_{e,i}\})
\label{projectv-id2}\eeq

which shows that correlations of the $\Psi_v$ wavefunction get
transferred to the $\Psi_e$ wavefunction upon ``projection''. Thus,
the zeroes of the bosonic wavefunction for the
pseudovortices  induce Laughlin-Jastrow
factors in the electronic wavefunction. This establishes the precise
connection between $\bR_v$ and the physical vortices.

Furthermore, it is well-known that wavefunctions with $2s$
Laughlin-Jastrow zeroes attached to the electrons have very nearly the
best energy for generic repulsive electronic interactions at
$\nu=p/(2sp+1)$. Thus, the $\Psi_v$ which produces these correlations
upon ``projection'', namely the $\nu={1\over2s}$ bosonic Laughlin
wavefunction (Eqn. (\ref{half-bos-laugh})), must therefore be very
nearly the best choice for $\Psi_v$.

\section{Details of the Hartree-Fock calculation}

Turning to the HF calculation of gaps we will once again use the {\em
preferred} density\cite{rest} in $H$ as we have in the past:
\beq 
\barro \to
\bar{\bar{\rho}}^p= \barro -c^2 \barchi 
\eeq 

This should make no difference in any exact calculation of gaps, which
can be seen as follows: First note that when we expand out $H^p =
H(\bar{\bar{\rho}}^p )$, there will be a $\barro \barro $ term, which
is there to begin with, a $\barchi \barchi $ term which contributes a
constant depending on our choice of $\Psi_v$ (equivalent to a choice
of ``gauge''), a constant which will drop out in the difference
between the ground state and a state with a widely separated
particle-hole pair, and finally a $\barro \barchi $ part whose
expectation value will vanish as along $\Psi_v$ is translationally
invariant. The reason we make the replacement $\barro \to \barro^p$ is
as in our earlier papers: in the HF calculation, this choice builds in
Kohn's theorem\cite{kohn} (leading to ${\bar S}(q)\simeq
q^4$\cite{GMP}) as well as the correct charge and dipole
moment\cite{read2} of the CF at tree level, thus making it plausible
that the results do not suffer strong vertex corrections\cite{rest}.

It is now straightforward, though tedious, to compute the correction
to the transport gap from the terms induced in the effective
Hamiltonian. One follows the rules of standard first order
perturbation theory and takes the average of the perturbation in the
unperturbed states. In this context, note that the HF states of
$H_{00}$ are not necessarily HF states of $H_{00}^{eff}$. However,
this effect is of order $v(q)/\omega_0$, and will affect the gaps to
second order in $v(q)/\omega_0$. In order to evaluate the averages in the
HF states, one needs to do two two-dimensional integrations. While it
is possible to do these numerically for an arbitrary potential, we
have simplified the problem by choosing a particularly tractable form
for the potential
\beq
v(q)={2\pi e^2\over q} e^{-\Lam^2 q^2}={2\pi e^2\over q}e^{-\lambda^2 q^2 l^2}
\label{potential}\eeq
This potential enables us to perform all the integrations
analytically. The relevant integrals are tabulated in Appendix B. 

Let us briefly sketch how the process works. We first represent the
projected preferred density in terms of CF annihilation and creation
operators $(d,d^{\dagger})$ in the Landau gauge:
\eqa
&\psi_{CF}(\vrr)=\sum\limits_{n,X} \phi_{n,X} d_{n,X}\\
\phi_{n,X}=&Ce^{iXy/(l^*)^2} e^{-(x-X)^2/2(l^*)^2}H_n((x-X)/l^*)
\eea
where $C$ is a normalization constant, $l^*=l\sqrt{2ps+1}$ is the
magnetic length in the effective field, and the $H_n$ are the Hermite
polynomials. In terms of $d_{n,X}$ and $\dd_{n,X}$, the preferred
density looks like
\beq
\bbrho(\bq)=\sum\limits_{\{n_i\}X}e^{-iq_xX}\dd_{n_1,X-{Q_yl^*\over2}} d_{n_2,X+{Q_yl^*\over2}} \bbrho_{n_1n_2}(\vq)
\eeq
where the matrix element is defined by 

\eqa
&\bbrho_{n_1n_2}(\vq)=(-1)^{n_<+n_2}\sqrt{{n_<!\over n_>!}}e^{i(n_1-n_2)(\t_q-\pi/2)}\nonumber\\
& \bigg((cQ/\sqrt{2})^{|n_1-n_2|}e^{-c^2Q^2/4}L_{n_<}^{|n_1-n_2|}(c^2Q^2/2)\nonumber\\
&-c^2(Q/c\sqrt{2})^{|n_1-n_2|}e^{-Q^2/4c^2}L_{n_<}^{|n_1-n_2|}(Q^2/2c^2)\bigg)
\label{bbrho}
\eea

The two-body and three-body terms have to be treated differently. The
two-body terms all involve the matrix elements of
$\bbrho(\bq)\bbrho(-\bq)$ integrated against different measures. One
first computes the contribution to the gap due to
$\bbrho(\bq)\bbrho(-\bq)$. Recall that the transport gap for
$\nu=\third$ is the energy difference between the $n=0$ CF-LL and the
$n=1$ CF-LL. Thus, 
\eqa
\delta(q)=&<n=1|\bbrho(\bq)\bbrho(-\bq)|n=1>-\nonumber\\
&<n=0|\bbrho(\bq)\bbrho(-\bq)|n=0>
\eea

This quantity is easily found to be 
\eqa
\delta(q)=&\bigg(2-{c^2\over1-c^2}(ql)^2\bigg)e^{-c^2(ql)^2/(2(1-c^2))}\nonumber\\
&+\bigg(-4c^2+{2c^2\over1-c^2}(ql)^2\bigg)e^{-(c^2+c^{-2})(ql)^2/(4(1-c^2))}\nonumber\\
&+\bigg(2c^4-{c^2\over1-c^2}(ql)^2\bigg) e^{-(ql)^2/(2c^2(1-c^2))}\nonumber\\
&+(ql)^2(1-c^2)e^{-(1-c^2)(ql)^2/(4c^2)}
\eea

It is seen that the result contains only powers of $ql$ and gaussian
factors. With the assumed expression for the interaction,
Eqn. (\ref{potential}), the integrations can be carried out
analytically.

The contribution of the three-body term to the gap is somewhat more
involved. It can be seen that it involves the expectation value of the
operator 
\beq
:\bbrho(\bq_1+\bq_2):\bbrho(-\bq_1)\bbrho(-\bq_2)
\eeq 
in the $n=0$
and the $n=1$ CF-LL's. After some effort, the expectation value of
this operator in the state $|m>$ can be computed to be
\eqa
&<m|:\bbrho(\bq_1+\bq_2):\bbrho(-\bq_1)\bbrho(-\bq_2)|m> = &\nonumber\\
&e^{\ihalf
\bq_1\times\bq_2\lst2} \sum\limits_{n_1n_2} (1-N_F(n_1))(1-N_F(n_2))\times&\nonumber\\
&\bbrho_{mn_1}(\bq_1+\bq_2)\bbrho_{n_1n_2}(-\bq_1)\bbrho_{n_2m}(-\bq_2)&
\nonumber\\
&-e^{-\ihalf
\bq_1\times\bq_2\lst2} \sum\limits_{n_1n_2} (1-N_F(n_1))N_F(n_2)\times&\nonumber\\
&\bbrho_{mn_1}(\bq_1+\bq_2)\bbrho_{n_1n_2}(-\bq_2)\bbrho_{n_2m}(-\bq_1)&
 \nonumber\\
&-e^{-\ihalf
\bq_1\times\bq_2\lst2} \sum\limits_{n_1n_2} N_F(n_1)(1-N_F(n_2))\times&\nonumber\\
&\bbrho_{mn_1}(-\bq_1)\bbrho_{n_1n_2}(\bq_1+\bq_2)\bbrho_{n_2m}(-\bq_2)&
 \nonumber\\
&-e^{\ihalf
\bq_1\times\bq_2\lst2} \sum\limits_{n_1n_2}(1-N_F(n_1)) N_F(n_2)\times&\nonumber\\
&\bbrho_{mn_1}(-\bq_1)\bbrho_{n_1n_2}(-\bq_2)\bbrho_{n_2m}(\bq_1+\bq_2)&
 \nonumber\\
&+e^{-\ihalf
\bq_1\times\bq_2\lst2} \sum\limits_{n_1n_2} N_F(n_1)N_F(n_2)\times&\nonumber\\
&\bbrho_{mn_1}(-\bq_2)\bbrho_{n_1n_2}(-\bq_1)\bbrho_{n_2m}(\bq_1+\bq_2)&
 \nonumber\\
&-e^{\ihalf
\bq_1\times\bq_2\lst2} \sum\limits_{n_1n_2} N_F(n_1)(1-N_F(n_2))\times&\nonumber\\
&\bbrho_{mn_1}(-\bq_2)\bbrho_{n_1n_2}(\bq_1+\bq_2)\bbrho_{n_2m}(-\bq_1)&
\eea

Only the $n=0$ CF-LL is occupied, and the infinite sums over $n_1$ and
$n_2$ (where not truncated by $N_F$) can be carried out using the
following expression for the matrix elements
\beq
\bbrho_{n_1n_2}(bq)=<n_1|e^{-ic\bq\cdot\etab}-c^2e^{-i\bq\cdot\etab/c}|n_2>
\eeq 
and using the completeness of the set of states $|n>$. Suffice it to
say that the results of this calculation can also be expressed
entirely in terms of powers and gaussians. Appendix B describes an
integral that can be used to find both the two-body and three-body
contributions to the gap.

\section{Results}
\begin{figure}
\narrowtext
\epsfxsize=2.4in\epsfysize=2.4in
\hskip 0.3in\epsfbox{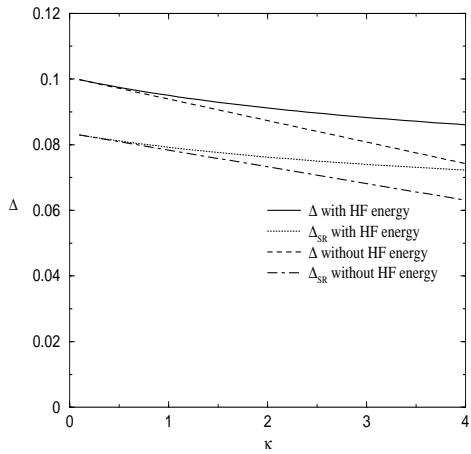}
\vskip 0.15in
\caption{Variation of the spin-polarized and spin-reversed gaps 
with LL-mixing for $\nu=\third$ at $\Lam=1.2l$.  The solid and dotted
lines have the electronic HF energy included (Eqn. (\ref{ehf}))
while the other two do not. All energies are in units of
$e^2/\varepsilon l$.
\label{fig1}}
\end{figure}

We have carried out this calculation for three cases, $\third$ and
${1\over5}$ (which are spin-polarized), and spin-singlet ${2\over5}$.
The common factor in all these cases is that only the $n=0$ CF-LL is
occupied. The only difference is in the value of $c$, which is
$c=\sqrt{2/3}$ for $\third$ and $\sqrt{4/5}$ for both ${1\over5}$ and
singlet ${2\over5}$. This leads us to the interesting fact in our
CF-HF approximation there is no difference between ${1\over5}$ and
singlet ${2\over5}$! In computing the effect of LL-mixing excitations
to all electronic LL's must be included in principle. In practice, we
found that the contributions drop rapidly for high $n'$. We found that
a cutoff of $n'_{max}=30$ was sufficient to capture all the
contributions to five-figure accuracy.

In Figure 1 we present the variation of the spin-polarized and
spin-reversed transport gaps for $\third$ as a function of $\kappa=
(e^2/\varepsilon l)/\omega_c$ for a ``thickness parameter''
$\Lam=1.2l$ (for which the HF gap in the limit of no LL-mixing has
appromimately the same value as for the true gap for Coulomb
interaction). The electronic HF energy (Eqn. (\ref{ehf})) has not been
added to the cyclotron energy in the dashed curve (which therefore
shows a linear dependence on $\kappa$), while it has been included in
the solid curve. It is seen the the gap decreases by a few percent for
realistic $\kappa$, but there is no dramatic effect.

\begin{figure}
\narrowtext
\epsfxsize=2.4in\epsfysize=2.4in
\hskip 0.3in\epsfbox{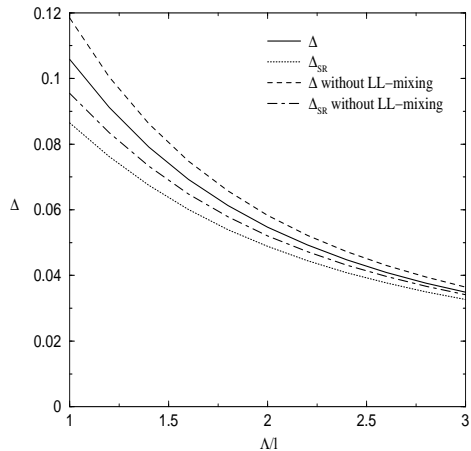}
\vskip 0.15in
\caption{Variation of the spin-polarized and spin-reversed gaps 
with $\Lam$ for $\nu=\third$ at $\kappa=2$.  For reference, the
results without LL-mixing are also presented. All energies are in
units of $e^2/\varepsilon l$.
\label{fig2}}
\end{figure}

In Figure 2 we show the variation of the gaps with $\Lam$ for fixed
$\kappa=1$, this time also plotting the result in the absence of
LL-mixing. It is seen that high values of $\Lam$ suppress Landau level
mixing. These results are in agreement with the previous fixed-phase
diffusion Monte Carlo (FPDMC) studies\cite{melik2} at a
semi-quantitative level. In general, our computation shows the gap to
be more robust than the FPDMC calculation. For instance, at
$\kappa=4$, which roughly corresponds to $r_s=10$, the FPDMC results
for the pure Coulomb interaction show a reduction of about 30\% in the
spin-polarized gap\cite{melik2} for $\third$. In contrast, our results
for $\lambda=1.2$, which has roughly the same gap in the absence of
LL-mixing, shows a reduction of only about 15\% (with the electronic
HF energy included in $H_{n'n'}$). On the other hand, this reduction is
about the same as the FPDMC results for a Coulomb interaction
corrected for sample thickness (thickness parameter
$\beta=1.5l$)\cite{melik2}. Our calculations for $\Lam=1.2l$ clearly
do include the strong suppression of the interaction at large
wavevectors, which is characteristic of the effects of sample
thickness. However, in order to carry out a detailed quantitative
comparison one would have to re-do our calculation with the same
interaction potential as was used in the FPDMC work. In the present
work, we have contented ourselves with a proof of principle,
concentrating on the analytically tractable potential of
Eqn. (\ref{potential}).

\begin{figure}
\narrowtext
\epsfxsize=2.4in\epsfysize=2.4in
\hskip 0.3in\epsfbox{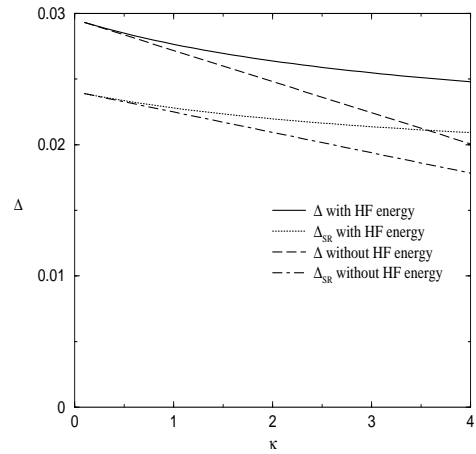}
\vskip 0.15in
\caption{Variation of the spin-polarized and spin-reversed gaps 
with LL-mixing for $\nu={1\over 5}$ and spin-singlet ${2\over5}$ at
$\Lam=0.8l$. Note that the spin-reversed gap applies only to ${1\over
5}$ while the spin-polarized gap applies to both ${1\over5}$ and
singlet ${2\over5}$.  The solid and dotted lines have the electronic HF
energy included (Eqn. (\ref{ehf})) while the other two do not. All
energies are in units of $e^2/\varepsilon l$.
\label{fig3}}
\end{figure}

Figure 3 displays the variation of the gaps with $\kappa$ for
${1\over5}$/singlet ${2\over5}$ (note that the spin-reversed gap is
only for ${1\over5}$), this time for $\Lam=1.6l$ (which again gives a
LLL gap roughly equal to that obtained from exact diagonalization for
the Coulomb interaction for $\nu={1\over5}$). 

Qualitatively, it is seen that the FQH states are very robust to
LL-mixing, at least as far as the transport gaps are concerned (which
has been pointed out before\cite{yoshioka,melik1,melik2}). While the
calculations which leave out the electronic HF energy in $H_{n'n'}$
show a linear dependence on $\kappa$, the ones which include the HF
energy show a much more physical behavior, with results that are even
more robust under LL-mixing.

\section{Conclusions}

The FQHE has been a fertile source of new ideas in the last two
decades. Much of the initial insight into the physics of the FQHE was
obtained by considering wavefunctions\cite{laugh,jain-cf}. However,
calculations of physical quantities based on the wavefunction approach
were forced to resort to computationally demanding techniques. The
advent of Chern-Simons field
theories\cite{cs-trans,gcs,zhk,read1,lopez,kalmeyer,hlr} of the FQHE
raised the possibility of approximate analytical schemes to calculate
the desired properties. Our Hamiltonian approach lies squarely in this
class of approaches.  Building on previous results, we were able to
obtain a LLL theory of CF's. Simple approximations such as HF in this
theory produce results\cite{rest} in very good agreement with those
obtained numerically on small systems, and with experiments in high
fields. However, as the field is lowered, the question of LL-mixing
becomes unavoidable, and must be accounted for theoretically. Previous
approaches to this issue have been based on exact
diagonalizations\cite{yoshioka} or the Quantum Monte Carlo
method\cite{price,melik1,melik2} on finite systems.

The approach to LL-mixing presented in this paper is analytical and
approximate. Despite the approximations, the results agree at a
semi-quantitative level with previous
results\cite{yoshioka,melik1,melik2}. For illustrative purposes, we
have chosen to compute the effect of LL-mixing on the transport
gap. In principle, one can calculate the effect of LL-mixing on {\it
any} physical property in this approach. Some, such as dynamical
response functions, or finite temperature spin polarization, would be
prohibitive to calculate numerically, but are easily computed in our
approach\cite{rest}. One interesting question concerns instabilities
of the FQH states to LL-mixing. This could be addressed in our
approach by applying a conserving approximation to calculate the
collective modes\cite{kada-baym,read3,rest}, and seeing whether any
instabilities develop in these modes as LL-mixing increases. We intend
to pursue this and other questions in the future.

We are grateful to the NSF for partial support under grants DMR0071611
(GM) and DMR0103639 (RS).

\section{Appendix A}

In this appendix we will examine how ``projection'' of a CFHF
wavefunction to a product form works in detail for $\nu=\third$. In
this case the ground state CFHF wavefunction is particularly simple,
being just the wavefunction of a fully filled CF-LL. In order to carry
out the projection we first have to reexpress this wavefunction in
terms of $\{\bR_{e,i},\bR_{v,i}\}$. Let us focus on the single-CF
states in the lowest CF-LL. These states have the property that they
are annihilated by the CF-cyclotron destruction operator. The
CF-cyclotron operators can be expressed as
\eqa
R_{CF,x}^{c}+iR_{CF,y}^{c}=&l\sqrt{{2\over1-c^2}}a_c\\
R_{CF,x}^{c}-iR_{CF,y}^{c}=&l\sqrt{{2\over1-c^2}}a_c^{\dagger}\\
R_{CF,x}^{g}+iR_{CF,y}^{g}=&l\sqrt{{2\over1-c^2}}a_g^{\dagger}\\
R_{CF,x}^{g}-iR_{CF,y}^{g}=&l\sqrt{{2\over1-c^2}}a_g
\eea
where
\eqa
\left[a_c,a_c^{\dagger}\right]=&\left[a_g,a_g^{\dagger}\right]=1\\
\left[a_c,a_g^{\dagger}\right]=&\left[a_g,a_c^{\dagger}\right]=0
\eea
We can now express these operators in terms of electron and
pseudovortex guiding center creation and annihilation operators, which
are defined by
\eqa
R_{e,x}+iR_{e,y}=&l\sqrt{2} A_e^{\dagger}={z_e\over 2}-2l^2{\bar \partial}_e\label{re+exp}\\
R_{e,x}-iR_{e,y}=&l\sqrt{2} A_e={{\bar z}_e\over 2}+2l^2{\partial}_e\label{re-exp}\\
R_{v,x}+iR_{v,y}=&{l\sqrt{2}\over c}A_v={{\bar z}_v\over 2}+{2l^2\over c^2} {\partial}_v\label{rv+exp}\\
R_{v,x}-iR_{v,y}=&{l\sqrt{2}\over c}A_v^{\dagger}={{z}_v\over 2}-{2l^2\over c^2} {\bar \partial}_v\label{rv-exp}
\eea
where $\partial_e,\ {\bar \partial}_e$ are shorthand for
${\partial\over\partial z_e}, \ {\partial\over\partial {\bar z}_e}$,
etc.  The resulting expressions are
\eqa
a_{g}=&{A_e-cA_v^{\dagger}\over\sqrt{1-c^2}}\\
a_{g}^{\dagger}=&{A_e^{\dagger}-cA_v\over\sqrt{1-c^2}}\\
a_c=&{A_v-cA_e^{\dagger}\over\sqrt{1-c^2}}\\
a_{c}^{\dagger}=&{A_v^{\dagger}-cA_c\over\sqrt{1-c^2}}
\eea

Consider now the CF state with CF-LL index and angular momentum
zero. This state will be labelled $|0,0>$ as satisfies
\eqa
a_c|0,0>=&(A_v-cA_e^{\dagger})|0,0>=0\label{aconvacc}\\
a_g|0,0>=&(A_e-cA_v^{\dagger})|0,0>=0\label{agonvacc}
\eea
Since all the available states are tensor products of LLL states of
electrons and pseudovortices we can express this state as
\beq
|0,0>=\sum\limits_{k_e,k_v=0}^{\infty} A(k_e,k_v) |k_e>_e\otimes |k_v>_v
\eeq
where 
\eqa
|k_e>_e=&{(A_e^{\dagger})^{k_e}\over\sqrt{k_e!}}|0>_e={1\over\sqrt{2\pi 2^{k_e} k_e!}} \bigg({z_e\over l}\bigg)^{k_e} e^{-|z_e|^2/4l^2}\\
|k_v>_v=&{(A_v^{\dagger})^{k_v}\over\sqrt{k_v!}}|0>_v= {1\over\sqrt{2\pi 2^{k_v} k_v!}} \bigg({cz_v\over l}\bigg)^{k_v} e^{-c^2|z_v|^2/4l^2}
\eea
are the LLL states of electrons and pseudovortices.

Applying the conditions of Eqs(\ref{aconvacc},\ref{agonvacc}) it can
easily be found that in $|0,0>$, one is forced to have  $k_e=k_v$ and 
\eqa
|0,0>=&\sqrt{1-c^2} \sum\limits_{k=0}^{\infty} c^k |k>_e\otimes|k>_v\\
&={c\sqrt{1-c^2}\over2\pi l^2}e^{-{|z_e|^2\over4l^2}-{c^2|z_v|^2\over4l^2}+{c^2z_ez_v\over2l^2}}
\eea

By similarly using the expression for $a_g^{\dagger}$ in terms of the
electron and pseudovortex coordinates and their derivatives one can
verify that
\beq
|0,m>={(a_g^{\dagger})^m\over\sqrt{m!}}|0,0> ={1\over\sqrt{m!}}\bigg(\sqrt{{1-c^2\over2}}\bigg)^m\bigg({z_e\over l}\bigg)^m|0,0>
\eeq
Thus, filling all the CF states in the CF-LLL leads to the wave function
\eqa
&\Psi_{\nu_{CF}=1}^{CF}=C\prod\limits_{i<j}(z_{e,i}-z_{e,j}) |0,0>\\
&=C\prod\limits_{i<j}(z_{e,i}-z_{e,j})e^{-{|z_e|^2\over4l^2}-{c^2|z_v|^2\over4l^2}+{c^2z_ez_v\over2l^2}}
\eea
where $C$ is a normalization constant. It can be seen that this is not
in the form of a product of functions of $z_e$ and $z_v$ alone.

Now consider ``projecting'' this wavefunction into a product form
against an arbitrary wavefunction $\Psi_v$ dependent on
$\{z_{v,i}\}$. Performing a few gaussian integrations one can easily
show that
\beq
\int d^2z_v f(\{{\bar z}_{v,i}\}) e^{-\sum\limits_{k} {c^2|z_{v,k}|^2\over2l^2} +\sum\limits_{k} {c^2z_{v,k}z_{e,k}\over2l^2}}=f(\{{z}_{e,i}\})
\label{projectv-id}\eeq

Now if one takes $\Psi_v$ to be the $\nu=\half$ bosonic Laughlin
wavefunction, it can be seen from eq(\ref{projectv-id}) that the
zeroes of the pseudovortex wavefunction are translated into
Laughlin-Jastrow factors in the electronic wavefunction. This
establishes the precise connection between correlations in the
pseudovortex coordinates and correlations in the projected electronic
wavefunction.

\section{Appendix B}

In this appendix we present an integral that can be used to find the
contributions of both the two-body and three-body terms. It is found
that the contributions to the gap in both cases can be expressed as a
product of powers of $(q_1l)^2$, $(q_2l)^2$, or $(\bq_1+\bq_2)^2l^2$,
and gaussians of the same arguments. This has to be integrated with a
complicated measure in $\bq_1$ and $\bq_2$.  Calling $q_1l=y_1$ and
$q_2l=y_2$, and using the specially chosen form of the interaction, it
is easy to see that all the integrals can be reduced to special cases
of the following integral

\eqa
&Q_n(\beta,\a_1,\a_2,\a_3,m_1,m_2,m_3)=\int\limits_{0}^{2\pi} {d\t_1d\t_2\over(2\pi)^2} \int\limits_{0}^{\infty} dy_1dy_2 \nonumber\\
&e^{-y_1^2(\half+\lambda^2+\a_1)} e^{-y_2^2(\half+\lambda^2+\a_2)} e^{-\a_3 y_1y_2\cos(\t_1-\t_2)} \nonumber\\
&{(-1)^n\over n(n!)} \bigg({y_1y_2\over2}\bigg)^n y_1^{m_1} y_2^{m_2} e^{im_3(\t_2-\t_1)} e^{\ihalf l^2(1+2\beta)\bq_1\times\bq_2}
\eea

One can perform the angular integration by choosing $\t=\t_2-\t_1$,
and using the identity

\beq
\int\limits_{0}^{2\pi} {d\t\over2\pi} e^{in\t} e^{ae^{i\t}+be^{-i\t}}=\bigg({b\over a}\bigg)^n I_{|n|}(2\sqrt{ab})
\eeq
where $I_n$ are the Bessel functions with imaginary argument.
 
This identity holds even for negative and/or complex values of $a$ and
$b$, which for our integral attain the values
$a={1\over4}+{\beta\over2}-{\a_3\over2}$ and
$b=-{1\over4}-{\beta\over2}-{\a_3\over2}$. 

For the radial integrals one uses the following two
formulas\cite{gradshteyn} (in which $J_{\nu}$ are the Bessel
functions, $\Phi$ is the confluent hypergeometric function, and $F$ is
the hypergeometric function)
\eqa
&\int\limits_{0}^{\infty} x^{\mu} e^{-\a x^2}J_{\nu}(\beta x)=&\nonumber\\
&{\beta^{\nu} \Gamma({\nu+\mu+1\over2})\over2^{\nu+1}\a^{{\nu+\mu+1\over2}}\Gamma(\nu+1)} &\Phi({\nu+\mu+1\over2};\nu+1;-\beta^2/4\a)\\
&\int\limits_{0}^{\infty} dt e^{-st} t^{b-1} \Phi(a;c;kt)&=\Gamma(b)s^{-b}F(a,b;c;k/s)\nonumber\\
&& \ \ \ [|s|>|k|]\nonumber\\
&=\Gamma(b)(s-k)^{-b} F(c-a,b&;c;k/(k-s))\nonumber\\
&& \ \ \ [|s-k|>|k|]
\eea

to obtain, for the case $ab>0$
\eqa
&Q_n(\beta,\a_1,\a_2,\a_3,m_1,m_2,m_3)={(-1)^n\over n} 2^{-1+(m_1+m_2)/2}\times\nonumber\\
&(2b)^{m_3+n} {\Gamma(n+(m_1+m_3+1)/2)\Gamma(n+(m_2+m_3+1)/2)\over n! (n+m_3)!} \times\nonumber\\
&{F(n+(m_1+m_3+1)/2,n+(m_2+m_3+1)/2;n+m_3+1;z)\over(1+2\lambda^2+2\a_1)^{n+(m_1+m_3+1)/2}(1+2\lambda^2+\a_2)^{n+(m_2+m_3+1)/2}}\nonumber\\
&
\eea
where 
\beq
z={4ab\over(1+2\lambda^2+2\a_1)(1+2\lambda^2+\a_2)}
\eeq

Similarly, for the case $ab<0$ we obtain
\eqa
&Q_n(\beta,\a_1,\a_2,\a_3,m_1,m_2,m_3)={(-1)^n\over n}2^{m_2-1}(2b)^{m_3+n}\times\nonumber\\
&(-4ab)^{-n-(m_2+m_3+1)/2} (\half+\lambda^2+\a_1)^{(m_2-m_1)/2}\times\nonumber\\
&{\Gamma(n+(m_1+m_3+1)/2)\Gamma(n+(m_2+m_3+1)/2)\over n! (n+m_3)!}z^{n+(m_2+m_3+1)/2} \times\nonumber\\
&F({m_3-m_1+1\over2},n+{m_2+m_3+1\over2};n+m_3+1;z)\nonumber\\
&
\eea
where
\beq
z={-4ab\over-4ab+(1+2\lambda^2+2\a_1)(1+2\lambda^2+\a_2)}
\eeq


\begin{thebibliography}{99}
\bibitem{fqhe-ex} D.Tsui, H.Stromer and A.Gossard, Phys. Rev. Lett.
{\bf 48},
1559, (1982).
\bibitem{perspectives} For a broad survey of the quantum Hall effects, see, 
{\it ``Perspectives in Quantum Hall Effects''}, 
S.Das Sarma and A.Pinczuk, Editors (Wiley Interscience, New York 1997). 
\bibitem{laugh} R.B.Laughlin,  Phys. Rev. Lett. {\bf 50},
1395, (1983)
\bibitem{jain-cf} J.K.Jain, Phys. Rev. Lett.  {\bf 63}, 199, (1989);
Phys. Rev.
{\bf B 41}, 7653 (1990); Science {\bf 266}, 1199 (1994).
\bibitem{jain-cf-review} J.K.Jain and R.K.Kamilla, in Chapter 1 of, {\it    
``Composite
Fermions''}, Olle Heinonen, Editor (World Scietific, Teaneck, NJ, 1998).
\bibitem{yoshioka} D. Yoshioka, J. Phys. Soc. Jpn. {\bf 55}, 885 (1986).
\bibitem{price} R. Price, P. M. Platzman, and S. He, \prl {\bf 70}, 
339 (1993); R. Price and S. Das Sarma, \prb {\bf 54}, 8033 (1996).
\bibitem{melik1} V. Melik-Alaverdian and N. E. Bonesteel, 
\prb {\bf 52}, R17032 (1995). 
\bibitem{melik2} V. Melik-Alaverdian, N. E. Bonesteel, 
and G. Ortiz, \prl {\bf 79}, 5286 (1997). 
\bibitem{us1} R.Shankar and G.Murthy, Phys. Rev. Lett. {\bf 79},
4437, (1997).
\bibitem{us2} G.Murthy and R.Shankar, in {\it ``Composite
Fermions''}, Olle Heinonen, Editor, (World Scientific, Teaneck, NJ,
1998).
\bibitem{aftermath} R. Shankar, \prl \ {\bf 83} 2382 (1999). 
\bibitem{rest} G. Murthy, { Jour. Phys. Cond. Mat.}\ {\bf 12}, 10543 (2000); 
R. Shankar, \prb \ {\bf 63} 085322 (2001); G. Murthy, \prb \ {\bf 60},
 13702 (1999); G. Murthy, \prb {\bf 64}, 195310 (2001).
\bibitem{dh} D.H.Lee, \prl {\bf 80}, 4745 (1998). \label{lee}
\bibitem{kohn} W.Kohn, Phys. Rev {\bf 123}, 1242 (1961).
\bibitem{GMP} S.M.Girvin, A.H.MacDonald and P.Platzman, Phys. Rev.
B
{\bf 33},
2481, (1986).
\bibitem{read2} N.Read, Semi. Sci. Tech. {\bf 9}, 1859 (1994);
Surf. Sci., {\bf 361/362}, 7 (1996).
\bibitem{cs-trans} J.M.Leinaas and J.Myrheim, Nuovo Cimento {\bf 37B}, 1 (1977). 
\bibitem{gcs} S.M.Girvin, in Chapter 9 of, {\it  ``The Quantum Hall
Effect''}, Edited by R.E.Prange and S.M Girvin, Springer-Verlag, 1990;
S.M.Grivin and A.H.MacDonald, Phys. Rev. Lett. {\bf 58}, 1252 (1987).
\bibitem{zhk}S.-C.Zhang, H.Hansson and S.A.Kivelson, Phys. Rev. Lett.
{\bf 62}, 82, (1989); D.-H.Lee and S.-C.Zhang, Phys. Rev. Lett. {\bf
66}, 1220 (1991); S.-C.Zhang, Int. J. Mod. Phys., {\bf B6}, 25
(1992).
\bibitem{read1} N.Read, Phys. Rev. Lett., {\bf 62}, 86 (1989).
\bibitem{lopez} A.Lopez and E.Fradkin, Phys. Rev. B {\bf  44}, 5246
(1991),
{\em ibid} {\bf 47}, 7080, (1993), Phys. Rev. Lett.  {\bf 69}, 2126
(1992).
\bibitem{kalmeyer} V.Kalmeyer and S.-C.Zhang, Phys. Rev. B {\bf 46},
9889 (1992).
\bibitem{hlr} B.I.Halperin, P.A.Lee and N.Read, Phys. Rev. B {\bf
47}, 7312 (1993).
\bibitem{kada-baym} G.Baym and L.P.Kadanoff, 
Phys. Rev. {\bf 124}, 287 (1961); 
L. P. Kadanoff and G. Baym, {\it ``Quantum Statistical Mechanics''}, Addison-Wesley, Reading, MA, 1989. 
\bibitem{read3} N.Read, \prb {\bf 58}, 16262 (1998).
\bibitem{gradshteyn} I. S. Gradshteyn and I. M. Ryzhik, 
{\em Table of Integrals, Series, and Products}, Fifth Edition, 
A. Jeffrey, Editor, Academic Press, New York, 1994. 

\end{thebibliography}
\end{document}